\documentclass[5p,times,authoryear]{elsarticle}

\usepackage{ecrc}

\volume{00}

\firstpage{1}

\journalname{Astronomy and Computing}

\runauth{O. Creaner, \& T. Carozzi}

\jid{ASCOM}

\jnltitlelogo{Astronomy and Computing}

\usepackage{amssymb}

\usepackage[figuresright]{rotating}

\usepackage[table]{xcolor}

\usepackage{makecell}

\usepackage{textcomp}

\usepackage[at]{easylist}

\usepackage{hyperref}

\begin{document}

\begin{frontmatter}



\dochead{\textcopyright{} 2019. This manuscript version is made available under the CC-BY-NC-ND 4.0 license \url{http://creativecommons.org/licenses/by-nc-nd/4.0/}}

\title{beamModelTester: software framework for testing radio telescope beams}


\author[DIAS]{Ois\'{i}n Creaner\corref{cor1}}
\cortext[cor1]{Corresponding author}
\ead{creanero@cp.dias.ie}

\author[OSO]{Tobia D. Carozzi}
\ead{tobia.carozzi@chalmers.se}

\address[DIAS]{Dublin Institute for Advanced Studies, 31 Fitzwilliam Place, Dublin 2, Ireland}

\address[OSO]{Department of Space, Earth and Environment, Chalmers University of Technology, Onsala Space Observatory, 439 92 Onsala, Sweden}

\begin{abstract}
The flux, polarimetric and spectral response of phased array radio telescopes with no moving parts such as LOFAR is known to vary considerably with orientation of the source to the receivers.  Calibration models exist for this dependency such as those that are used in the LOFAR pipeline. Presented here is a system for comparing the predicted outputs from any given model with the results of an observation. In this paper, a sample observation of a bright source, Cassiopeia A, is used to demonstrate the software in operation, by providing an observation and a model of that observation which can be compared with one another.  The package presented here is flexible to allow it to be used with other models and sources. The system operates by first calculating the predictions of the model and the results of an observation of linear fluxes and Stokes parameters separately. The model and observed values are then joined using the variables common to both, time and frequency.  Normalisation and RFI excision are carried out and the differences between the prediction and the observation are calculated.   A wide selection of 2-, 3- and 4-dimensional plots are generated to illustrate the dependence of the model and the observation as well as the difference between them on independent parameters time, frequency, altitude and azimuth. Thus, beamModelTester provides a framework by which it is possible to calibrate and propose refinements to models and to compare models with one another.  

\end{abstract}


\end{frontmatter}


\section{Introduction}
\label{Introduction}
This paper presents a software package, \texttt{beamModelTester} \citep{beamModelTesterGithub}, which is designed to enable the evaluation and comparison of models of the beams of radio telescopes.  This is especially valuable for flux calibration of radio telescopes with no moving parts such as the LOw Frequency ARray (LOFAR).  The antennas in such telescopes are fixed in position and receive signals from all directions at once, in contrast to traditional, mechanically-slewed radio telescopes where the dish(es) and antenna(s) are moved to point at the target source. Instead, the signals from each of the stationary antennas are combined using software and electronic components \citep{butcher2004lofar}.   LOFAR has, since its development, been referred to as a ``software telescope'' \citep{butcher2004lofar}.  Software is needed to perform all operations on the telescope, and software-based models are used extensively to calibrate the outputs \citep{butcher2004lofar}. 

\begin{figure}[!htb]
         \center{\includegraphics[width=.47\textwidth]
         {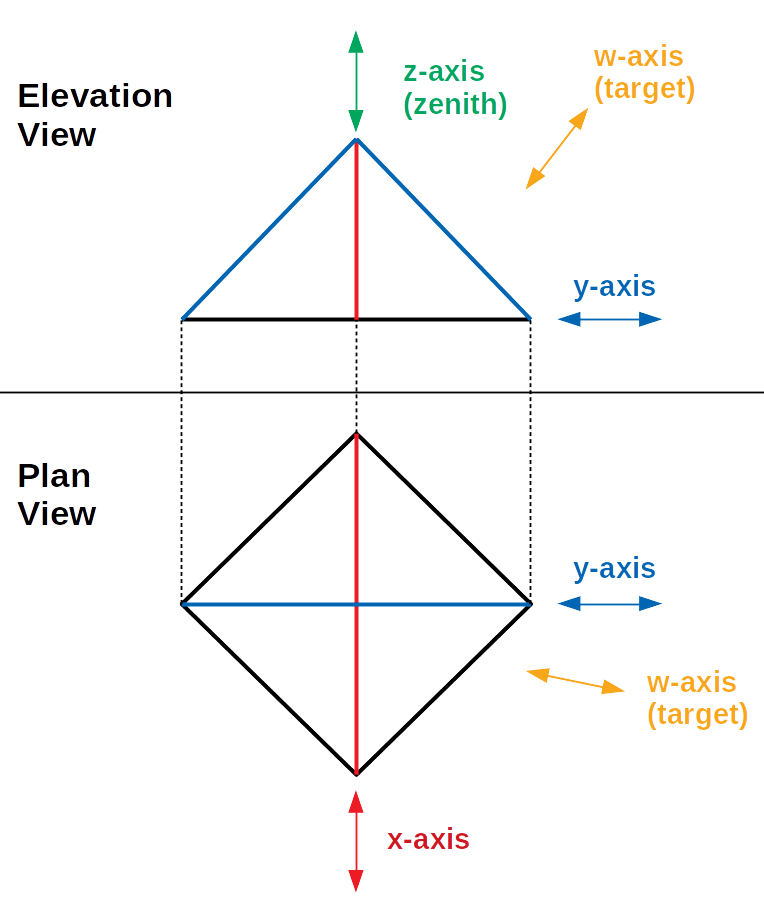}}
         \caption{\label{fig:LOFAR_PLAN}A schematic representation of a single LBA element of a LOFAR station in Elevation and Plan view.  The element consists of a ground plane (represented in black) and two identical antennas (represented in red and blue) supported at the centre of the element and orthogonal to one another.  Each antenna defines an axis along its length (\textit{x}-axis in red, \textit{y}-axis in blue) and the axis orthogonal to both defines a third axis \textit{z} (shown in green).  An arbitrary target in direction \textit{w} can be selected such that  $\angle$\textit{wx} and $\angle$\textit{wy} can independently vary between 0\textdegree and 90\textdegree .  Note that in the Elevation view, the \textit{x}-axis is viewed end-on, while in the Plan view, the \textit{z}-axis is viewed end-on.}
       \end{figure}

In Figure \ref{fig:LOFAR_PLAN}, a schematic of the configuration of a single Low Band Antenna (LBA) element can be seen as an example of such an instrument.  The element consists of two antennas which are aligned orthogonally to one another on axes labelled \textit{x} and \textit{y}.  These two axes are orthogonal to the \textit{z}-axis which,  in the case of LOFAR, is fixed to point straight up from the ground towards zenith.  Each LOFAR station consists of a number of such elements, each with the same alignment of \textit{xyz}-axes.  Other similar instruments, including LOFAR High Band Antenna (HBA) stations, have similar overall structure, but may differ in details.

The telescope is electronically ``steered'' towards a target in a reference direction \textit{w}\footnote{this is the same \textit{w} from which the \textit{uvw}-coordinates are calculated for imaging studies} by combining signals from each element with appropriate delays. Since the \textit{w}-axis can be chosen arbitrarily, the angle between it and the \textit{x}-axis can be seen to vary independently to the angle between it and the \textit{y}-axis.  

The relative geometric orientation of the inbound radiation from a source to that of a dipole antenna leads to a variation in the voltage response of that antenna.  A simple element of this variation comes from the fact that E-M waves are transverse, and because of this, the sensitivity is maximised when the \textit{w}-axis is orthogonal to the antenna, and minimised when the \textit{w}-axis is parallel with the antenna.  Because for LOFAR, the \textit{z}-axis is fixed towards zenith, the sensitivity is maximised for a source at zenith while at lower altitudes, this sensitivity can be reduced substantially as the angle between the source and the antenna decreases.    Since all antennas in a station are aligned on the same axes, this antenna-level effect could be expected to impact the station-level response as well \citep{dreamBeamPresentation}.

As an object rotates about the celestial pole over the course of a sidereal day, there is a separate variation for each of the two orthogonal antennas. This is because the angle between the source and the linear antennas can widen and narrow separately, changing the response of each antenna.  This can lead to an apparent polarisation in the detected signal.  Again, this contrasts with mechanically steered telescopes where the axes of the telescope can be rotated to follow the source and maintain maximum sensitivity in all directions.  Additional complexities for modelling this variation occur as signals interact with multiple antennas in a station, and atmospheric effects are taken into consideration \citep{di2019electromagnetic}.

Thus, there is a variation with respect to altitude (used in this paper in the sense of angle above the horizon) and azimuth (angle East/West of the Northern meridian) which must be modelled.  For the study of flux- or polarisation-variability of objects, it is essential to correct for this instrumental variation by means of calibration.  In the case of LOFAR, these calibrations are carried out based on an analytical model, such as that initially developed by \citet{hamaker2011mathematical} which is integrated into the data processing software (e.g. Default Pre-Processing Pipeline (DPPP) \citep{DP3github}).  These models are discussed in more detail in Section \ref{Existing}.

\texttt{beamModelTester} is designed to compare the predictions of these models with outputs from real observations to provide an assessment of the quality of these models.  Throughout this paper, plots are shown that provide examples of this software system in operation, which include automatically generated titles and axis labels.  These automatically generated plots are shown with a black outline.

\section{Sample Reference Measurement}
\label{Initial}
In order to provide a demonstration of the system in operation, it was necessary to produce an observation and a model which could be compared with one another using the software presented here.  The requirement for the observation was that it be of a bright source which was circumpolar from the available observatory, ideally passing close to zenith.  The source was required to be point-like at the resolution of the instrument that was used. 

       \begin{figure}[!htb]
         \center{\fbox{\includegraphics[width=.47\textwidth]{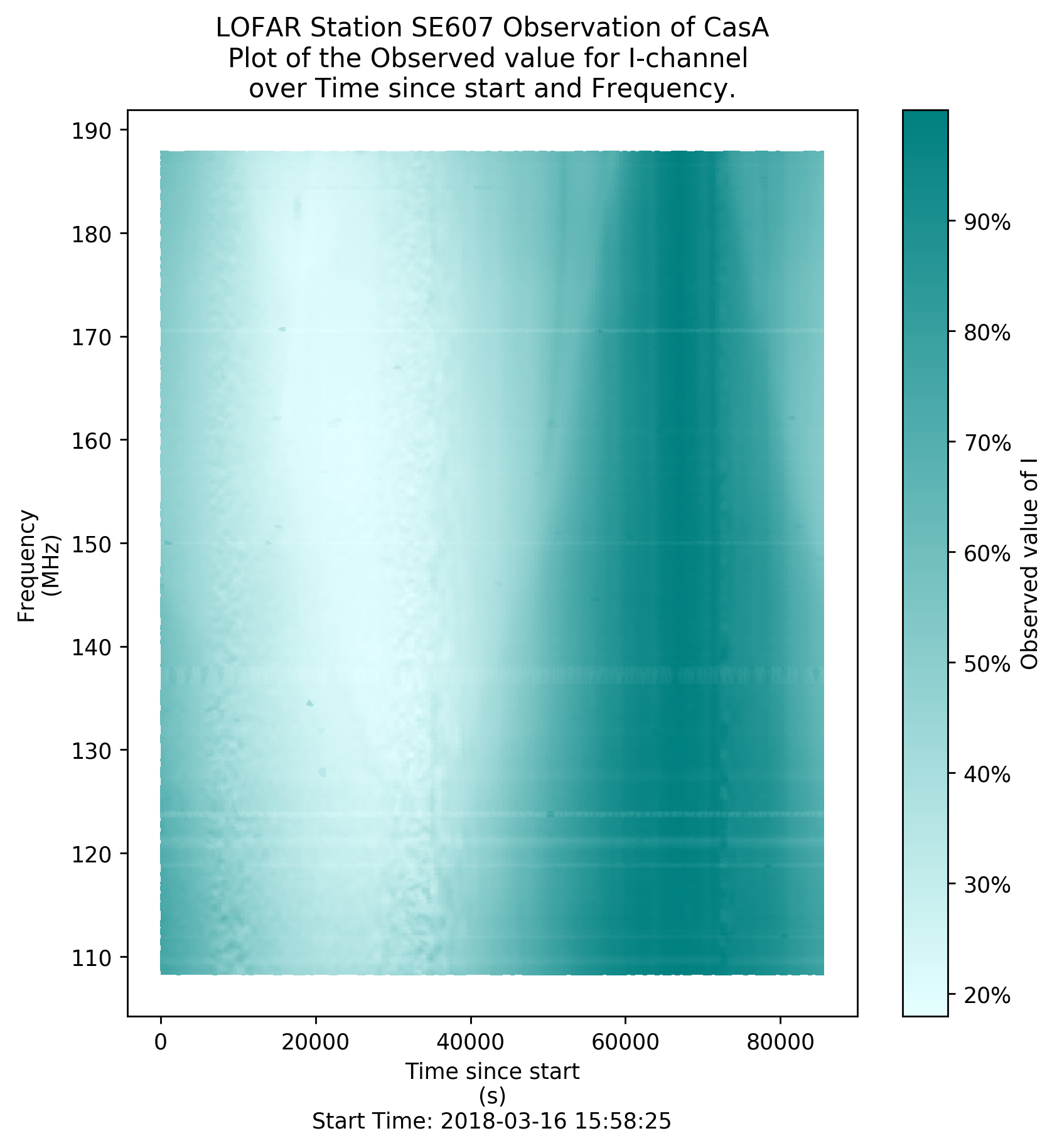}}}
  \caption{\label{fig:SE607_IvTimevFreq}A plot of observed flux (Stokes I) from CasA against Time and Frequency as observed with LOFAR Station SE607 HBA over 12 hours. This plot has been trimmed to remove RFI-dominated frequencies and normalised across frequencies using the maximum method.  Plot (including title and axis labels) automatically generated using \texttt{beamModelTester}.}
\end{figure}
     
A 24-hour observation of the radio source Cassiopeia A (CasA) taken on 16th-17th March 2018 from LOFAR HBA station SE607 at Onsala Space Observatory, Sweden is used a sample source to demonstrate this software in operation throughout this paper.  The timescale of variation in flux from CasA is known to be very much longer than the 24 hours over which the observation took place \citep{helmboldt2009evolution}.  

The station is located at Latitude: 57\textdegree 23$^\prime$ 35$^{\prime\prime}$ N, Longitude: 11\textdegree 55$^\prime$ 04$^{\prime\prime}$ E.  The HBA consists of 96 HBA tiles, each of which consists of 16 pairs of orthogonal antennas.  Each tile feeds into two Receiver Control Units (RCUs), one for each polarisation.  The maximum separation between antennas in the HBA field is 60m.  The maximum frequency in this observation is 200MHz, giving a minimum wavelength of 1.5m.  Therefore, the maximum resolution is $\sim$1.4\textdegree. By comparison, the angular size of CasA at these wavelengths is $\sim$5$^\prime$, \citep{arias2018low} and thus can be treated as a point source for the purposes of this experiment. 
       
The observation consisted of recording a series of array covariance matrices (ACMs) also known as ``crosslets'' or ``visibilities,'' at a rate of 1 per second \citep{virtanen2012station}. These were stored in Array Covariance Cubes (ACC Files).  Each ACC File consists of 512 ACMs, one for each subband in the HBA range.  A 5 second interval between the end of one ACC recording and the start of the next gave an observation cadence of one complete observation of the frequency range of the HBA every 519 seconds. These ACMs were converted into beamformed observations of a point source using \texttt{iLiSA} \citep{iLiSAGithub} as discussed in Section \ref{Solution}. Because this 519s observation cadence is short compared with the 24-hour runtime of the observation, observations in a single ACC file are treated as taking place simultaneously.

Figure \ref{fig:SE607_IvTimevFreq} demonstrates the variation of the apparent flux (Stokes I: calculated as the sum of the \textit{x}- and \textit{y}-axis fluxes) from CasA over the course of this observation.  This variation with respect to orientation is not fully consistent with respect to frequency.  While the general trend remains for intensity in each subband to reach a maximum at about the same time and a minimum at about the same time, the curve between these points has noticeably different shapes at different frequencies.  The apparent variation in CasA is believed to be induced by its changing orientation with respect to the antennas.

 \begin{figure}[!htb]
         \center{\includegraphics[width=.47\textwidth]
         {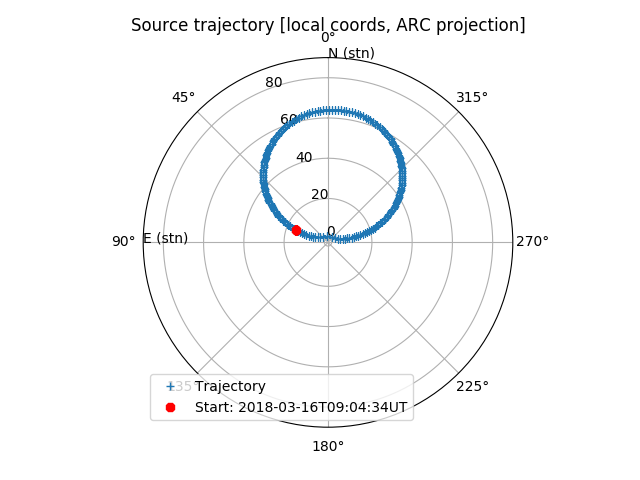}}
         \caption{\label{fig:CasA_path}Path of CasA on the sky as observed from LOFAR station SE607 over 24 hours, passing just North of zenith at a maximum altitude of 88.58\textdegree.  Plot generated using \texttt{dreamBeam}. This output displays the position in zenith angle (90\textdegree - altitude) and azimuth.}
       \end{figure}      
        
By taking the position of the station, the target and the time, it is possible to calculate the location of the target on the sky at any time as discussed in Section \ref{Solution}.  Figure \ref{fig:CasA_path} shows the path taken by CasA on the sky during the 24-hour observation. From SE607, CasA reaches a maximum altitude of 88.58\textdegree{} (i.e. very close to zenith) and the lowest altitude of 26.24\textdegree. 

\begin{figure}[!htb]
         \center{\fbox{\includegraphics[width=.47\textwidth]
         {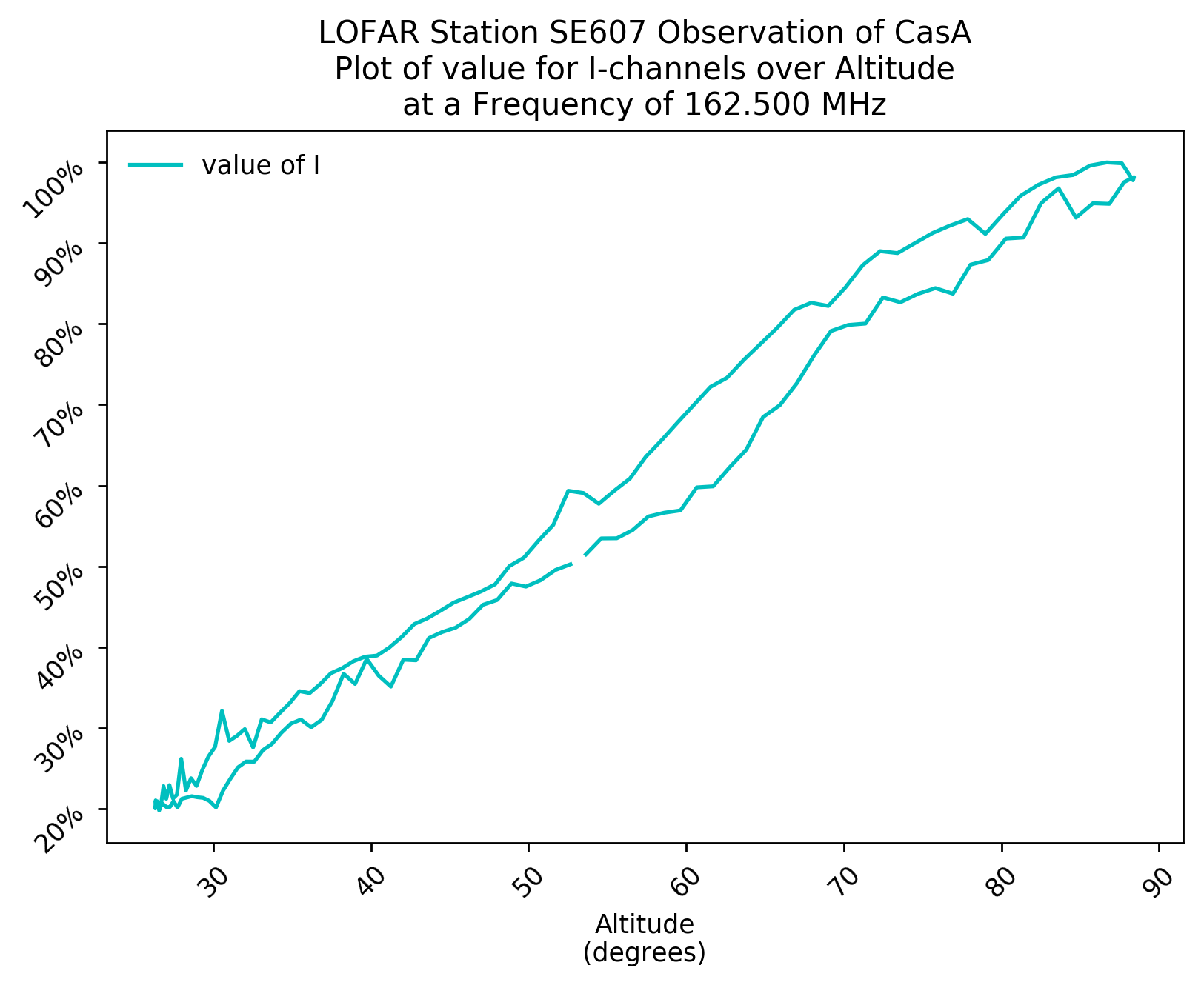}}}
         \caption{\label{fig:IvAlt}A plot of observed flux (Stokes I) from CasA against Altitude as observed with LOFAR Station SE607 HBA at 162.5MHz over 24 hours.  Plot (including title and axis labels) automatically generated using \texttt{beamModelTester}.}
       \end{figure}

An example of the variation in flux with respect to altitude can be seen in Figure \ref{fig:IvAlt}.  For this plot, the altitude and azimuth are calculated for each of the time steps and the former is used as an axis for plotting.  As is shown, the flux varied by a factor of up to $\sim$5 from near-zenith to its lowest altitude at the sample frequency of 162.5MHz.  

\begin{figure}[!htb]
         \center{\fbox{\includegraphics[width=.47\textwidth]
         {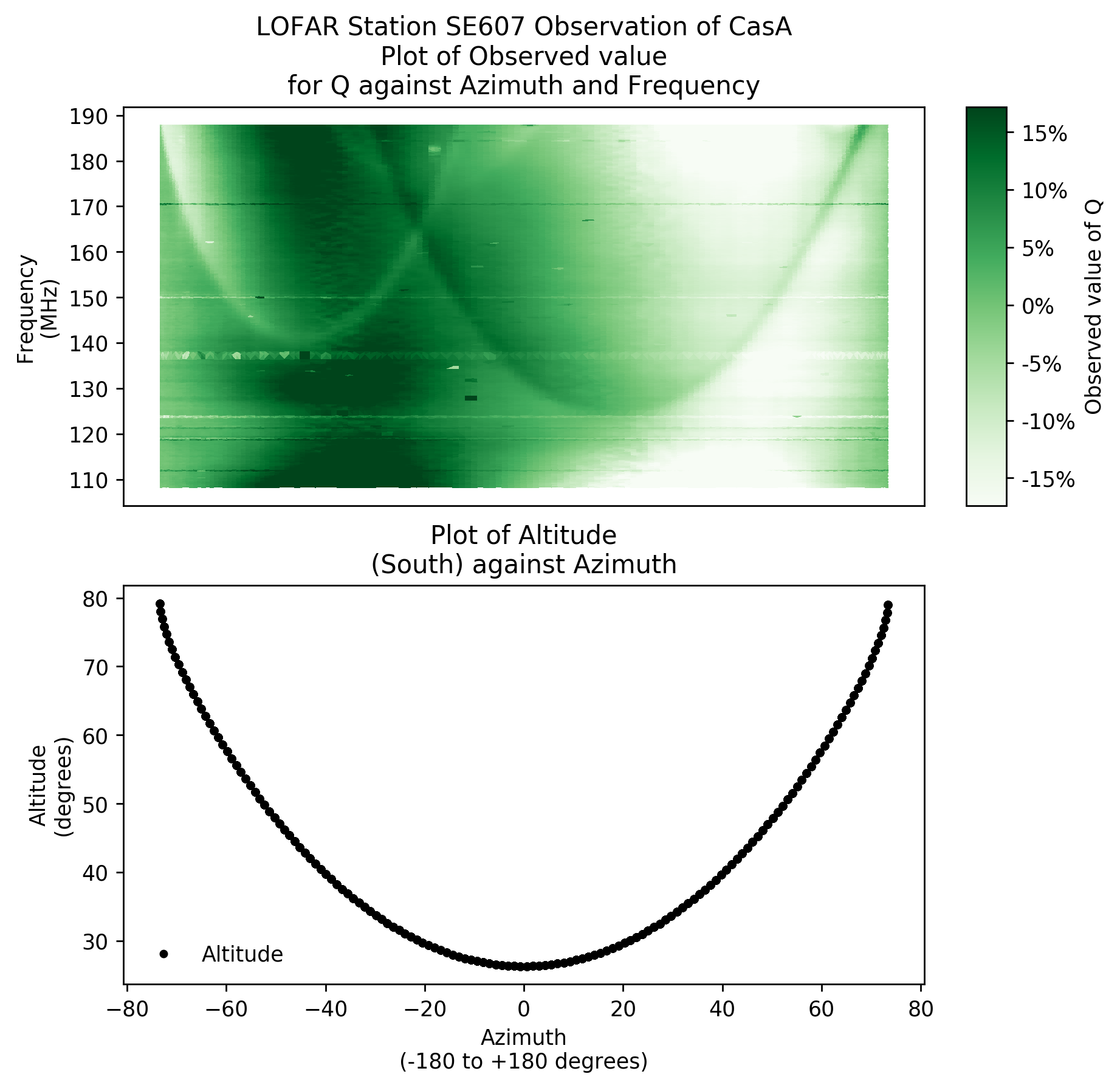}}}
  \caption{\label{fig:SE607_QvAzvFreq}(top) A plot of observed linear polarisation (Stokes Q) against Azimuth and Frequency. (bottom) A plot of Altitude against Azimuth at the corresponding times of this observation to indicate the variation in two independent variables.  This plot has been trimmed to remove RFI-dominated frequencies.  Paired plot (including title and axis labels) automatically generated using \texttt{beamModelTester}.}
  
       \end{figure}

In addition, since antenna orientation defines linear polarisation channels, and the relative orientation between source and antenna causes variation in antenna response, it follows that as altitude changes lead to a change in both x- and y-antenna response, azimuth changes lead to a different response for each of the linear antennas. This means that an apparent linear polarisation (Stokes Q: calculated as the difference between the \textit{x}- and \textit{y}-axis fluxes) will arise in a system with linear antennas which do not move as the source azimuth to the detector changes with the rotation of the earth as shown in Figure \ref{fig:SE607_QvAzvFreq}.
      
Figure \ref{fig:SE607_QvAzvFreq} demonstrates  the major trend in apparent polarisation over azimuth can be seen in form of a peak at azimuth $\sim$45\textdegree W and the trough at $\sim$45\textdegree E for most frequencies, representing source orientations where the source is aligned approximately along the length of one antenna and perpendicular to the other.  Additional complexity in the structure can be seen which is discussed briefly in Section \ref{Results}.

Since response in flux and polarisation can clearly be seen to vary against frequency, time, altitude and azimuth, it follows that any model developed must be able to incorporate each of those variables.

A model of this observation was generated using \texttt{dreamBeam} \citep{dreamBeamGithub}. \texttt{dreamBeam} is  a piece of software designed to implement existing models of telescope performance.  \texttt{dreamBeam} produces predictions of the flux from an ideal point-like source at given sky coordinates as observed from a given observing location.  None of the models currently implemented in \texttt{dreamBeam} incorporate sky models.  As used in this system it outputs the predictions of the model to a \texttt{CSV} file.

\section{Existing Models}
\label{Existing}

A mathematical framework developed by J. P. Hamaker provides a mathematical-physical basis for the description of the variation in the response of a LOFAR station based on a mix of analytical and numerical simulations \citep{hamaker2011mathematical}. The coefficients for this framework were implemented by M. Arts. This framework forms part of DPPP \citep{DP3github,imagingCookbook}. An implementation of this pipeline is used in \texttt{dreamBeam} to calculate predicted values for observations \citep{dreamBeamGithub}.

This model is known to have some limitations and reservations which must be addressed \citep{hamaker2011mathematical}.  Firstly, the simulation is based on an extension of a model of a single antenna under ideal conditions i.e. away from electromagnetic obstructions \citep{hamaker2011mathematical}.  In order to produce an array such as is used in LOFAR, such obstructions are all around in the form of the other antennas in the array \citep{asad2015polarization,di2019electromagnetic}.  Secondly, the model, by design, is based on the open-circuit voltage, i.e. it does not include impedance matching \citep{hamaker2011mathematical}.

The model, therefore, is known to be incomplete in its predictions.  Quantitative assessment of the degree to which this model diverges from practical observation are therefore required to enable users of the system to have confidence in its predictions.  

In addition, ongoing work on refined or other alternative models has been discussed by \citet{asad2015polarization,asad2016polarization} and \citet{di2019electromagnetic} amongst others.  To compare the performance of these models with the model used in the LOFAR Pipeline to determine which should be used it is necessary to develop a robust structure to provide a figure of merit for the performance of the model.

Finally, it is hoped that by comparing models with observations and with one another, it might be possible to guide further refinements to the models by highlighting areas of the spectrum or sky in which the performance of the model is deficient.

\section{Requirements for testing system}
\label{Requirements}

It is therefore apparent that any such system of modelling would require a robust system to test and evaluate its performance and such a system is presented here. The key features of a testing system are that it be modular, flexible and user-friendly.

In this context, modularity is essential under two categories.  Data modularity is the requirement that the system must be able to handle inputs from disparate systems - model data from each of a variety of models, observation data from a variety of telescopes - where the data structure cannot be guaranteed to be uniform between different systems.  Modules capable of loading data from each possible source are therefore necessary.  Functional modularity is the design requirement that mandates that components of the system be able to work independently from one another, which allows independent features to be added to the system in response to user demand without compromising existing functionality.

Flexibility is the requirement to provide immediate, tailored outputs in a variety of use cases, to enable the examination of a variety of parameters of the models to be tested.  Further, the capture of additional use cases for related objectives, such as automated production of dynamic spectra can be achieved through flexible design.

A user-friendly system is responsive to user demands, has clear input mechanisms and gives outputs which are clear and instantly recognisable.  By focussing on human-centred design, the presentation of output results enables robust analysis for follow up studies.

\section{Software Solution Design}
\label{Solution}

\begin{figure}
        \center{\includegraphics[width=.47\textwidth]
        {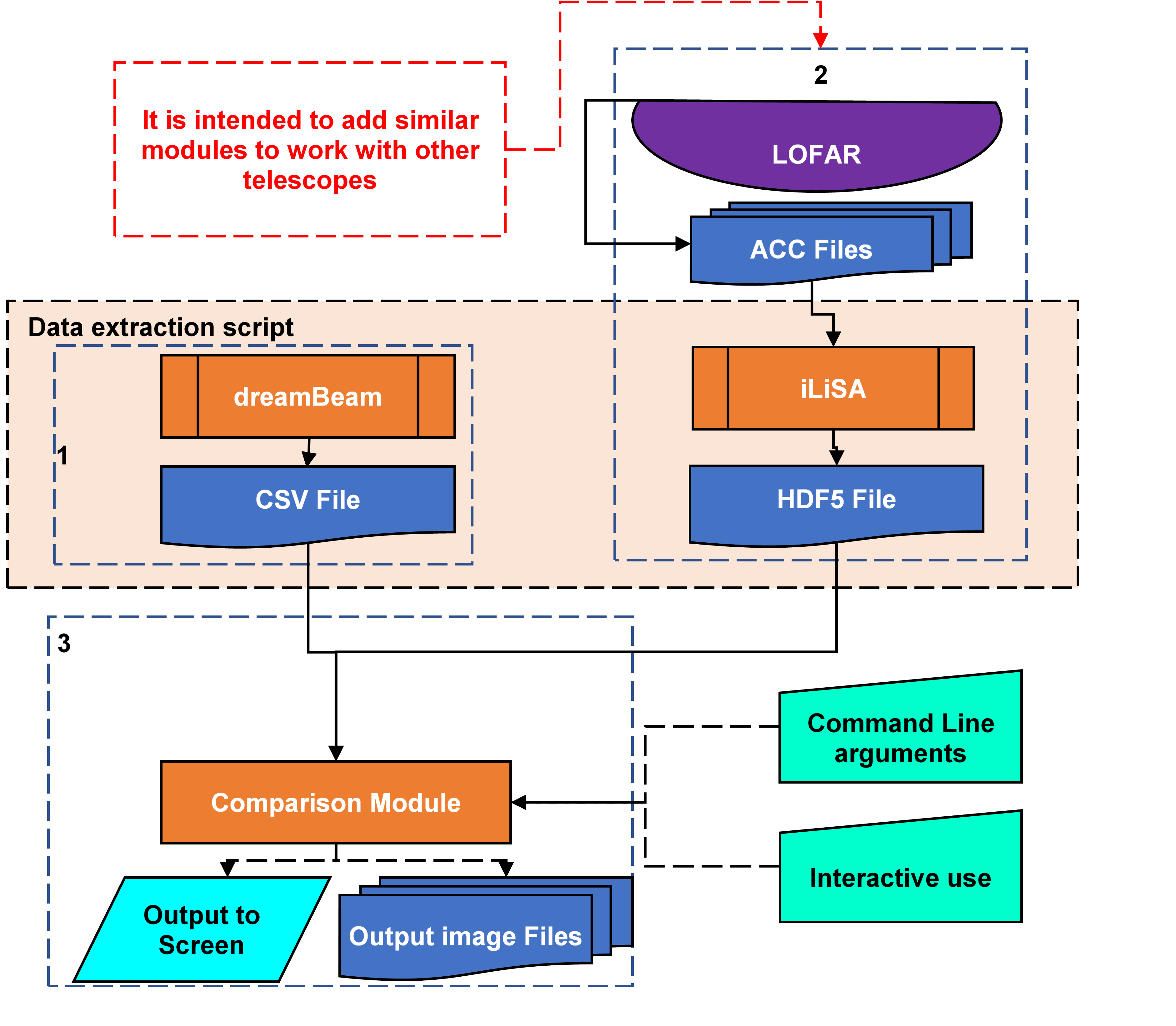}}
        \caption{\label{fig:flowchart}Overall design of solution.  \texttt{dreamBeam} \citep{dreamBeamGithub} provides model data of the predicted variation of a target over the course of an observation in a \texttt{CSV} file. \texttt{iLiSA}  \citep{iLiSAGithub} converts \texttt{ACC} files output from the telescope into a \texttt{HDF5} file with fluxes at given times and frequencies for a given target.  The comparison module of \texttt{beamModelTester} \citep{beamModelTesterGithub} brings these together in memory for processing into charts.}
      \end{figure}

The solution presented here consists of three main modular components, each designed to ``plug-in'' seamlessly to the others, and which can be replaced with equivalent components for future modules and applications.  As shown in Figure \ref{fig:flowchart}, the three major elements are model data calculation and input (currently provided by \texttt{dreamBeam}, \citet{dreamBeamGithub}), observation data recording and transformation (provided by \texttt{iLiSA}, \citet{iLiSAGithub}) and a module to combine and compare data from each of the input components.

Model data is provided through \texttt{dreamBeam}, a multi-purpose system which implements the beam models, such as the existing model as discussed above and, as used in this suite, produces outputs in the form of text which can be stored as a \texttt{CSV} file \citep{dreamBeamGithub,dreamBeamPresentation}.  Inputs to \texttt{dreamBeam} specify a target, model, observing location and time for the observation.  The output is a normalised prediction of variation over time (and frequency) for an uncontaminated point source in Jones Matrix form.  

Observed data is currently provided through \texttt{iLiSA} (International LOFAR in Stand Alone mode).  This system has multiple elements, but the feature used here is the ability to use cross-correlation files - specifically Array Covariance Cubes (\texttt{ACC} files) - to generate beamformed fluxes for a given target within the beam of the telescopes used.  This calculation is carried out without assumptions regarding a sky model and thus does not include demixing of other sources.  It also operates under the assumption that the target is point-like at the resolution of the single station.  Outputs from \texttt{iLiSA} in this mode consist of \texttt{HDF5} files containing the observed times, frequencies and linear channel fluxes \textit{xx}, \textit{xy} and \textit{yy} \citep{iLiSAGithub}.

The testing system brings these two together by providing flexible plug-ins for the data to \texttt{beamModelTester}.  The reading functions of this suite of software read the data from each of the sources into a \texttt{Pandas dataframe} \citep{mckinney-proc-scipy-2010}.  Linear fluxes and Stokes parameters are then calculated from the Jones Matrix data. Data from the two sources are joined using common variables date/time and frequency to allow to enable comparisons between the model and the observed data.  Throughout this system, all times are assumed to follow the same standard, usually UTC.  Conversion between local time and UTC is possible by adding an appropriate time offset as part of the user input. 

These programs are tied together using two wrapper scripts.  These scripts transfer the variables and data from one program to another.  Note that separate scripts are provided to access data in raw form and data that has already been processed once.  Data processing from raw data takes time, but the intermediate data formats (\texttt{CSV} and \texttt{HDF5} files as shown in Figure \ref{fig:flowchart}) are much faster to load into memory than to generate from the source format \texttt{ACC} files.

The Horizontal coordinates Altitude and Azimuth are calculated from the station geographic coordinates and the celestial coordinates of the target for each timestamp in the observation using the \texttt{coord.transform\textunderscore to}  method of \texttt{Astropy} \citep{astropy:2013,astropy:2018}.  An East-West version (as opposed to 0-360\textdegree{}) of Azimuth is calculated by subtracting 360\textdegree{} from all Azimuths over 180\textdegree.  A rotated set of coordinates called \textit{Station} coordinates are calculated from the Horizontal coordinates by means of the \texttt{antennafieldlib} method of \texttt{iLiSA} \citep{iLiSAGithub}.  \textit{Station} coordinates are rotated such that the \textit{x}- and  \textit{y}-axes of a given station are aligned with the ordinal (NE, SE, SW and NW) points of the \textit{Station} coordinate system.  

This creates two sets of variables that can be compared, the independent variables and the dependent ones.  The independent variables are time, frequency, altitude and azimuth, the latter of which can be in either conventional horizontal or station relative coordinates. The dependent variables are a set of different formulations of measures of flux and polarisation.  These consist of the linear antenna fluxes (xx and yy), the cross-correlation antenna component (xy) and the Stokes parameters (I, Q, U, V.)

The system is designed for use from the command line via input parameters, via a text-based menu interface or using a graphical user interface (GUI).  Each of the interfaces enables the user to select the dependent and independent variable(s) to plot, as well as to select between various representation approaches and output data types and locations.  The code, together with detailed instructions and tutorials on its use are available from \citet{beamModelTesterGithub}.

\section{Comparison Mechanisms}
\label{Mechanisms}
Comparisons between observation and model values can be made by three main approaches, Direct Comparison, Difference Plots and Figures of Merit.

\paragraph{Direct Comparison}
Plots which show both model and observed data and enable the user to compare the two while retaining clear view of the original data.
\begin{itemize}
\item \textit{Separate 3-d colour and/or contour plots} of the variation of a dependent variable against the independent variables can be generated for the model and observation data to be used in side-by-side comparisons for 3-variable data (e.g. flux against frequencies and time, as shown in Figure \ref{fig:SE607_IvTimevFreq})
\item \textit{Separate 3-d colour plots with context plot} of the variation of a dependent variable against the independent variables can be generated for the model and observation data to be used in side-by-side comparisons for 4-variable data (e.g. flux against frequencies and azimuth with a context plot showing altitude against azimuth as shown in Figure \ref{fig:SE607_QvAzvFreq})
\item \textit{Line plots} of model and observed data can be generated to allow for direct comparison for 2-variable data (e.g. flux against altitude at a fixed frequency as shown in Figure \ref{fig:IvAlt}) which may be overlaid upon one another or shown side-by-side.
\item \textit{Animated line plots} of model and observed data where one independent variable is assigned to time, and another to the x-axis (e.g. showing how the apparent spectrum varies over time.)  Again, these can be overlaid or placed side-by-side.
\end{itemize}
\paragraph{Difference Plots}
Plots of the difference between the model and the observation.  These plots can be displayed with or without the original data, and can be plotted in the same manner as the direct plots above. Differences can be calculated in the forms
\begin{itemize}
\item \textit{Subtraction} (model-observation) This approach determines the difference between the model and the observation by subtracting the value of the observation from that of the model.  In order for this to be a sensible approach, the data must be in the same units and appropriately normalised.  The output is viable even in cases where one or both values are zero or negative
\item \textit{Division} (model/observation) This approach determines the difference between the model and the observation by reference to the observation.  The variation in this value indicates a variation in the ratio between the two measurements, and can be used without normalisation.  Outputs can be difficult to interpret if there are negative values for some measurements, and is undefined if the observed value, or both values are zero.
\item \textit{Inverse division} (observation/model) This approach determines the difference between the model and the observation by reference to the model. The variation in this value indicates a variation in the ratio between the two measurements, and can be used without normalisation.  Outputs can be difficult to interpret if there are negative values for some measurements, and is undefined if the model, or both values are zero.
\end{itemize}
\paragraph{Figures of Merit}
Calculations and plot of figures of merit for the variation between model and observation, and how those figures vary across independent variables.  The figures of merit used are Root Mean Square Error (RMSE) and Pearson's Correlation.
\begin{itemize}
\item \textit{RMSE}  This figure of merit is a measure of the average separation of the model value from the observed value.  It remains valid even if some values are negative or zero. In order for this to be a sensible approach, the data must be in the same units and appropriately normalised.
\item \textit{Pearson's Correlation}  This is a measure of how similar the pattern of variation in the observation matches that in the model.  This figure of merit can be used without normalisation.
\end{itemize}

\section{Cropping and Normalisation}
\label{Cropping}

In addition to the above mentioned plotting options, there are two other factors which must be addressed when comparing observations with models: elimination of Radio Frequency Interference (RFI), and ensuring comparable units.

RFI can be a major issue, especially for instruments such as LOFAR HBA which operates in the same frequency bands as artificial sources such as radio communications \citep{offringa2013lofar}.   Two approaches to eliminate RFI exist within the current software solution.  Additionally, the modular nature of the software enables the development of more sophisticated RFI filtering options in the future if needed.

The first method of removing RFI is a frequency filter.  This filter allows for only specified frequencies to be plotted in the output data.  Known frequencies/subbands where RFI is strong (e.g. known Air Traffic Control channels) can be removed from a list of subband start-points and input either manually or as a file containing a list of viable frequencies.

The second method to eliminate RFI is cropping.  The user can specify thresholds above which data points are eliminated.  This cropping can be operated globally, or on a per-frequency or per-time basis, can the threshold can be set using a percentile, or using a multiple of the median or mean values in the data.  Cropping allows for automated elimination of outliers but suffers in that it can lead to the elimination of real data as well as RFI-driven outliers.

To ensure that the model inputs are comparable to the inputs from observations, the data must be normalized.  In this system, various options for normalisation are provided, depending on the requirements of the user.  The normalisation system is modular, and can be extended with additional options in future versions of the software.  The current normalisation operations are Maximum-based normalisation and Fit-based normalisation, Maximum-based normalisation means to divide by the maximum value in each of the input columns in the dataset, before calculating the derived values.  Fit-based normalisation calculates the linear multiplication factor and constant offset that provides a least-square fit between the model and the observation, and applies these factors to the observation.

These normalisations can be applied to the data either overall, on a per-frequency or a per-time basis. In overall normalisation mode, the system applies the normalisation process to the whole dataset at once.  In frequency- and time-normalisation mode, the unique values of each of these independent variables are determined.  The normalisation operation is carried out on the data corresponding to each of these values.  Frequency mode enables comparison to be made effectively between observations which necessarily are dependent on the spectral response of the telescope, and models of those variations which do not account for the overall sensitivity curve.  Time-normalisation allows the user to examine variations in the shape of the detected spectrum as the target rotates about the pole.

\section{Results}
\label{Results}

This system has been fully implemented and is operational.  As such, it enables users of radio telescopes with no moving parts such as LOFAR to calibrate for a variety of different conditions and allows for examination of many issues that can arise due to the relative alignment of source and detector.  

\begin{table}[htb!]
\begin{tabular}{|p{1.45cm}|p{4.4cm}|p{1.65cm}|}
\hline
 \textbf{Variable}                          & \textbf{Description} & \textbf{Dependent?} \\ \hline \hline
\textbf{xx} & x-axis flux                        & Dependent                 \\ \hline
\textbf{xy} & x/y-axis cross-correlation*                       & Dependent                 \\ \hline
\textbf{yy} & y-axis flux                        & Dependent                 \\ \hline
\textbf{Stokes I} & xx+yy overall flux                        & Dependent                 \\ \hline
\textbf{Stokes Q} & xx-yy linear polarisation                        & Dependent                 \\ \hline
\textbf{Stokes U} & Real (xy) polarisation angle                        & Dependent                 \\ \hline
\textbf{Stokes V} & Im (xy) circular polarisation                        & Dependent                 \\ \hline
\textbf{Frequency} & Frequency of subband start                        & Independent                 \\ \hline
\textbf{Time} & Time of the start of ACC file                        &  Independent                 \\ \hline
\textbf{Altitude} & Angle of target above Horizon as viewed from the station                        & Independent                 \\ \hline
\textbf{Azimuth} & Angle of target East of North from the station                        &  Independent                 \\ \hline
\textbf{Azimuth (E/W)} & Angle of target East (+) or West (-) of North from the station                        &  Independent   \\  \hline
\textbf{Station Altitude} & Angle of target above Horizon in \textit{Station} coordinates                        & Independent                 \\ \hline
\textbf{Station Azimuth} & Angle of target East of North in \textit{Station} coordinates                        &  Independent                 \\ \hline
\textbf{Station Azimuth (E/W)} & Angle of target East (+) or West (-) of North in \textit{Station} coordinates   &  Independent \\  \hline   
\end{tabular}

 \caption{A list of variables that can be plotted, with an indication as to whether they are treated as a dependent or independent variable.  *xy is a complex value.  When plotting, the absolute value is plotted.}
 \label{table:variables} 
\end{table}

For a given observation, each of the set of variables representing flux and polarisation described in Section \ref{Solution} and summarised on Table \ref{table:variables} can be plotted against a set of independent variables:  altitude, azimuth or time on one axis and frequency on the other.  These plots can be generated as 3-d contour or colour plots, as animated plots or as individual frames  depending on the needs of the user.  Outputs can be stored in any suitable location, and a variety of file types are available, depending on the user and the software environment.  Data from source and model can be filtered, cropped and normalised together or separately using a variety of options for each parameter.   Plots can be overlaid or plotted separately as needed to allow for direct comparisons or to clear up graphical plots. These plots are summarised on Table \ref{table:plots}.

\begin{table}[htb!]
\begin{tabular}{|p{2.80cm}|p{3.45cm}|p{1.25cm}|}
\hline
 \textbf{Plot}                          & \makecell{\textbf{Variables} }& \textbf{Overlay?} \\ \hline \hline
\textbf{2-d line plot} & \makecell{1 Ind (x-axis)\\ N Deps (y-axis)} & Yes                \\ \hline
\textbf{3-d colour plot} &  \makecell{2 Inds (x \& y-axes) \\ 1 Dep (z-axis) } & No             \\ \hline
\textbf{3-d contour plot} &  \makecell{2 Inds (x \& y-axes) \\ 1 Dep (z-axis) } & No                \\ \hline
\textbf{Animated line plot} &  \makecell{2 Inds (x \& t-axes)\\  N Deps (y-axis) } & Yes                \\ \hline
\makecell[l]{\textbf{3-d colour plot} \\ \textbf{with context plot}} &  \makecell{3 Inds (x, y\textsubscript{1} \& y\textsubscript{2}-axes) \\ 1 Dep (z-axis) }& No                 \\ \hline
\end{tabular}

 \caption{The Plot column shows a list of the types of plots that can be generated.  The Variables column shows the types of variables (i.e. Independent/Dependent Variables) that can be plotted, based on Table \ref{table:variables} and the axis they are mapped to.  The Overlay column indicates whether multiple plots can be shown overlaid upon one another for plots of a given type. (e.g. as shown in Figure\ref{fig:SE607_yyvAlt})}
 \label{table:plots} 
\end{table}

       \begin{figure}[!htb]
         \center{\fbox{\includegraphics[width=.47\textwidth]
         {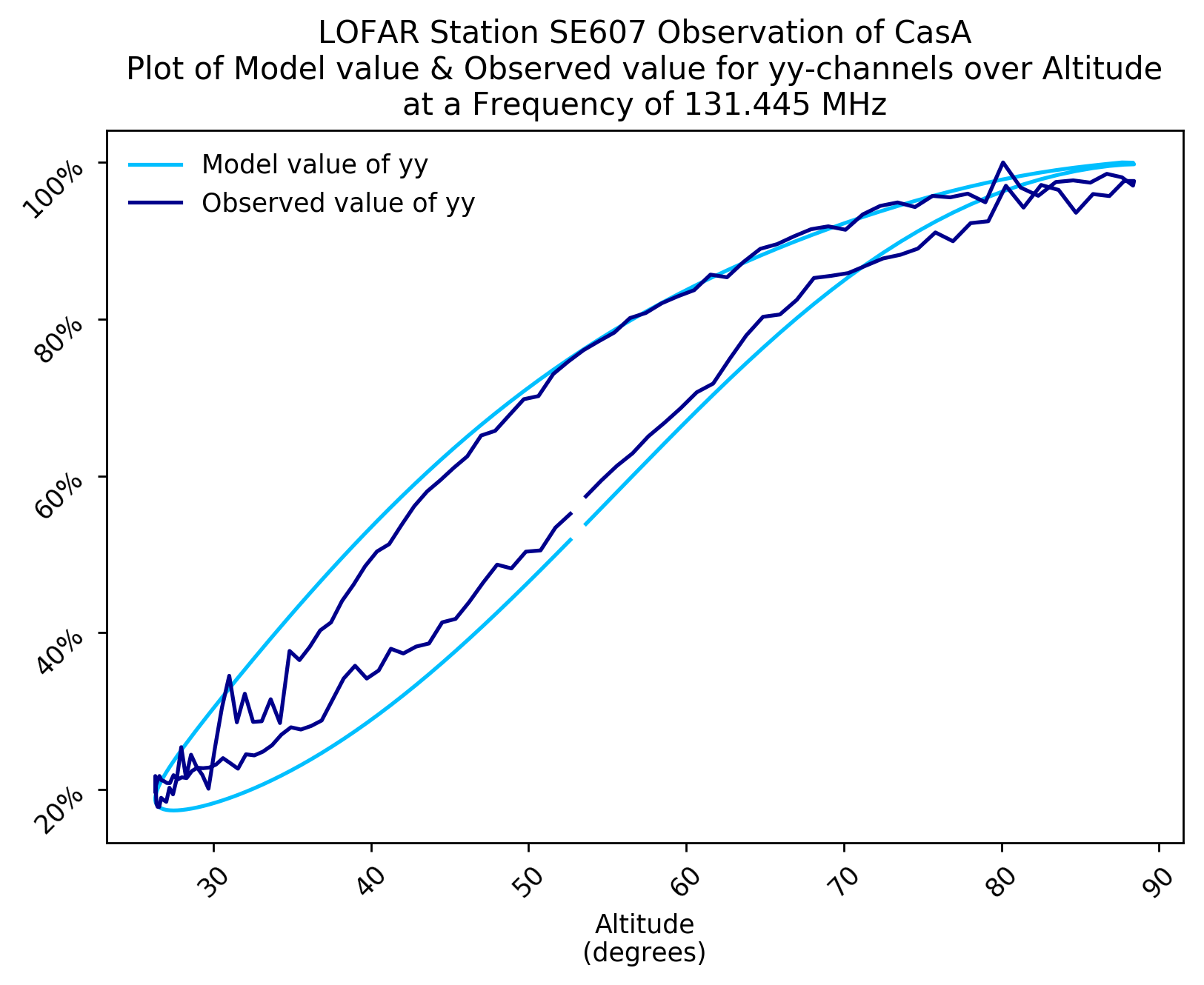}}}
  \caption{\label{fig:SE607_yyvAlt}A plot of observed and model flux in the \textit{y}-polarisation channel from CasA against Altitude as observed with LOFAR Station SE607 HBA at 131.445MHz over 24 hours. Illustration of an area of the spectrum where the model and the observation are in close agreement. At higher altitude, noise levels are greater than the model-source disagreement.  Plot (including title and axis labels) automatically generated using \texttt{beamModelTester}.}

\end{figure}

Taken together, these parameters allow for many thousands of possible combinations and permutations for a given set of input data depending in the use-case required, and it would be impossible to fully discuss them all here.  A number of sample observations which demonstrate the utility of the system are outlined below.

       \begin{figure}[!htb]
         \center{\fbox{\includegraphics[width=.47\textwidth]
         {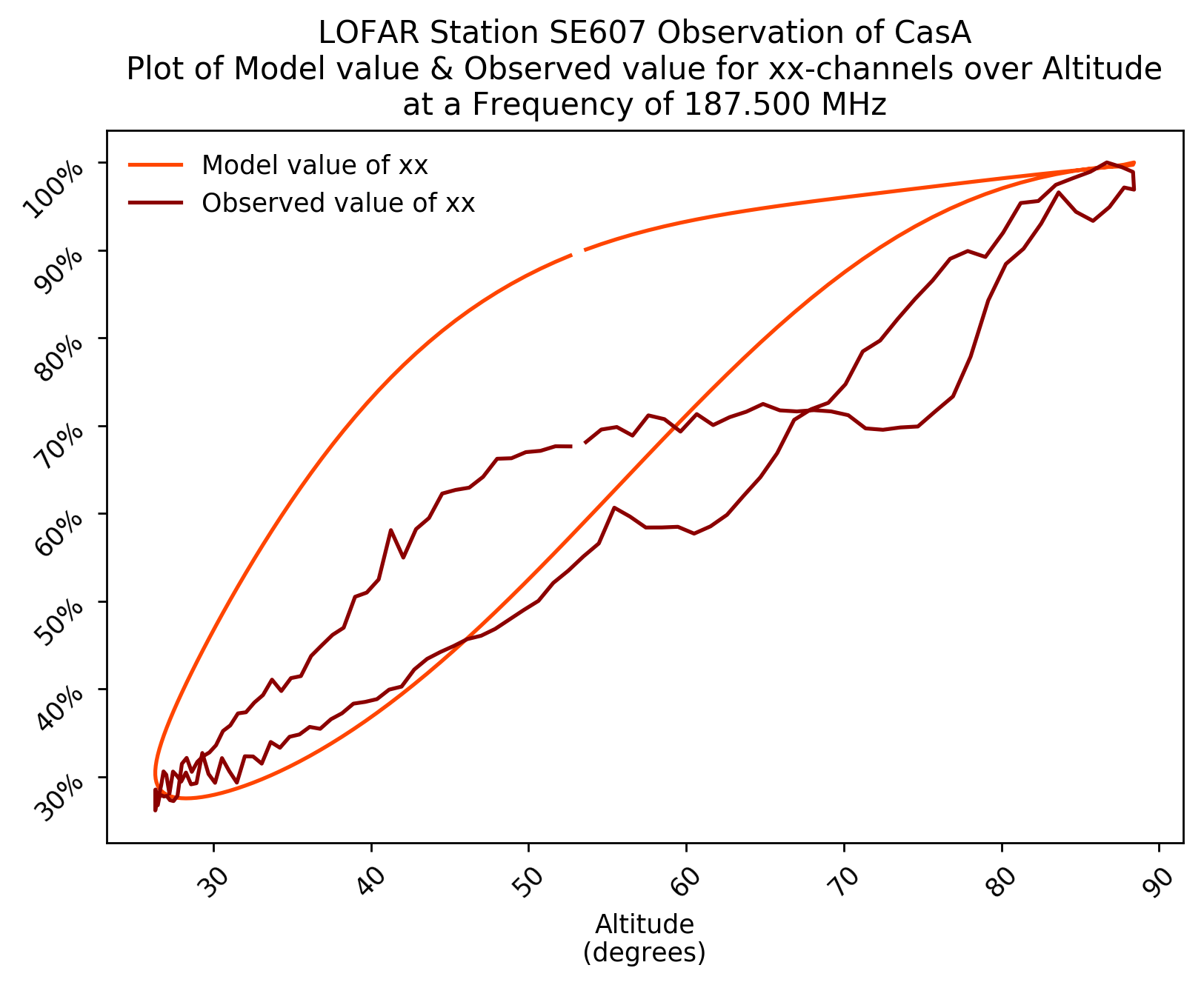}}}
  \caption{\label{fig:SE607_xxvAlt}A plot of observed and model flux in the \textit{x}-polarisation channel from CasA against Altitude as observed with LOFAR Station SE607 HBA at 187.500MHz over 24 hours. Illustration of an area of the spectrum where the model and observed data deviate strongly from one another.  Here, neither the general shape of the distribution nor the specific values are in agreement.  Plot (including title and axis labels) automatically generated using \texttt{beamModelTester}.}
       \end{figure}
       
Figures \ref{fig:SE607_yyvAlt} and \ref{fig:SE607_xxvAlt} each show the comparison between the \texttt{dreamBeam} output (labelled model) and the \texttt{iLiSA} output (labelled observed) as well as the differences between them.  Overlaid plots such as these, which are taken from frames from an animated plot of this comparison, allow for areas in the distribution where model and source diverge significantly to be observed, to enable modellers to understand which regions to focus on when refining their models.  For example, in Figure \ref{fig:SE607_yyvAlt}, especially at high altitudes, the model and observation are in close agreement, and thus the difference can be shown to be low.  On the contrary, in Figure \ref{fig:SE607_xxvAlt}, the shape and position of the observed curve is significantly different to the model.  This suggests a region of the radio spectrum in which the model should be modified to account for additional factors.

       \begin{figure}[!htb]
         \center{\fbox{\includegraphics[width=.47\textwidth]{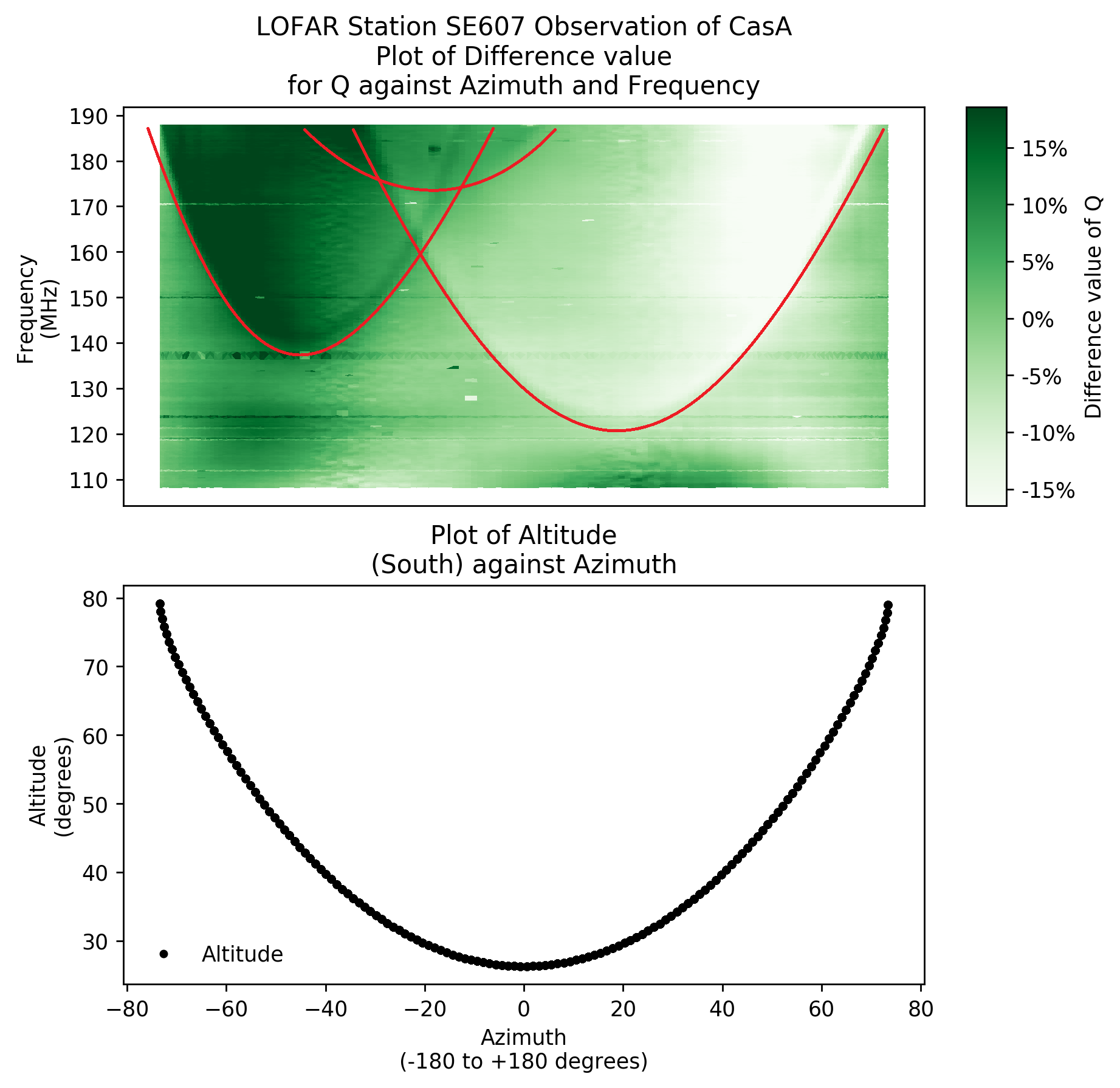}}}
  \caption{\label{fig:SE607_QvAzvFreq_diff}(top) A plot of the difference between observed and model of linear polarisation (Stokes Q) against Azimuth and Frequency. (bottom) A plot of Altitude against Azimuth at the corresponding times of this observation to indicate the variation in two independent variables.  Note the parabola-like contamination from sidelobe observations of other A-Team sources.  Paired plot (including title and axis labels) automatically generated using \texttt{beamModelTester}.  Red curves manually added to draw attention to patterns in the data.}
  
       \end{figure}

Plotting the data in three dimensions can allow for patterns to be noted in the distribution of divergences between the model and the observation.  As is shown in the upper plot of Figure \ref{fig:SE607_QvAzvFreq_diff}, there are a number of smooth curved shapes which have been highlighted. \citet{tasse2012lofar} suggests some explanations which can be applied to features such as these.  For example: many can be explained by the non-point-like behaviour of radio telescope beams. One explanation is that features like these can be caused by the side-lobes of the LOFAR beam shape covering another object in the so-called ``A-team'' set of sources.  

\begin{figure}[!htb]
         \center{\includegraphics[width=.47\textwidth]{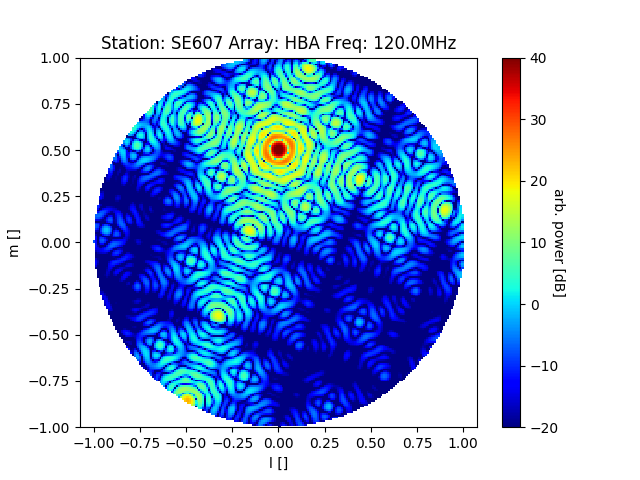}}
  \caption{\label{fig:beam_pattern_120}Orthographic projection model of the Array Factor of LOFAR HBA station SE607 at 120MHz. Note the extensive sidelobe patterns reaching all the way to the horizon.  Plot generated using \texttt{dreamBeam}.}
\end{figure}

\begin{figure}[!htb]
         \center{\includegraphics[width=.47\textwidth]{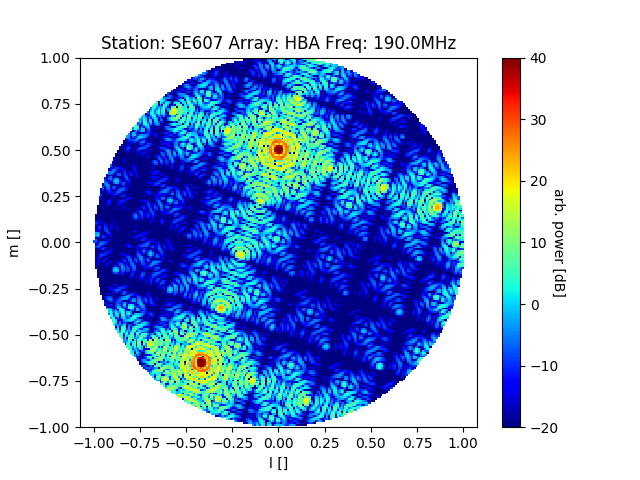}}
  \caption{\label{fig:beam_pattern_190}Orthographic projection model of the beampattern of LOFAR HBA station SE607 at 190MHz. Note the differences in the beampattern from that shown in Figure \ref{fig:beam_pattern_120}  In particular, note the narrower spacing of the fringes and the strong local maximum located close to the horizon.  Plot generated using \texttt{dreamBeam}.}
\end{figure}

In particular, LOFAR HBA observations are vulnerable to the presence of bright objects in the side lobes of the beam pattern, as the regular and even spacing of HBA antennas can lead to extreme sidelobe patterns such as that shown in Figure \ref{fig:beam_pattern_120}.  The shape of these patterns is dependent on the \textit{uv} spacing of the antenna elements, which will necessarily change as the w-axis changes to follow the observed object. Additionally, as the size, strength and spacing of these sidelobes is wavelength-dependent, cross-contamination will not necessarily occur at the same time, altitude or azimuth for all frequencies as illustrated by the differences between Figures \ref{fig:beam_pattern_120} and \ref{fig:beam_pattern_190}.  As a result, the additional contaminated flux can appear to move across wavelengths as the target source moves across the sky.  This is believed to produce the distinctive parabola-like structures shown in Figure \ref{fig:SE607_QvAzvFreq_diff}.
       
 
It is apparent from Figure \ref{fig:SE607_QvAzvFreq_diff} that the model output from the implementation of the Hamaker model in \texttt{dreamBeam} does not account for features such as these.  Should a model be available to account for these deviations, it can be tested with \texttt{beamModelTester} and a reduced difference between the model and the observation would be expected in the corresponding plots.

\section{Conclusions}
\label{Conclusions}

Use of this software system enables a user wishing to quantify the performance of a model of radio telescopes with no moving parts to robustly compare the model with a real observation.  By plotting the observation alongside the model and/or plotting the difference between the observation and model, the user can determine any areas where the model does not give an accurate representation of real observations.  This can be used to identify additional factors, such as cross-contamination by second (and more) sources in the side-lobes of a beam which must be accounted for in any attempt to calibrate the observation by means of a model.

The design of the software solution enables it to be extended to use additional inputs with minimal modification to the system.  Notably, if new models are developed which are intended to more closely describe expected observations, these can be either plugged in directly, if their output format matches an existing format, or a suitable plugin developed to import the data, which can then be compared against observation using the existing system.

Ongoing refinements to the system are envisaged to take the form of upgrades, rather than invalidating existing use of the system.  Planned upgrades include the ability to integrate multiple targets or observations from multiple stations to provide more complete sky coverage.

The software system has been successfully designed, implemented, and is available together with user and design documentation at the link below:
\url{www.github.com/creaneroDIAS/beamModelTester}

A large selection of sample outputs are available at the link below.
\url{www.github.com/creaneroDIAS/beamModelTester/blob/master/comparison_module/outputs.md}

\section*{Acknowledgements}
This publication has received funding from the European Union’s Horizon 2020 research and innovation programme under grant agreement No 730562 [RadioNet].

The software developed in this project makes use of the following software packages:
\begin{itemize}
\item \texttt{dreamBeam}: a Radio telescope beam modeling framework \citep{dreamBeamGithub}
\item \texttt{iLiSA}: international LOFAR in Stand-Alone mode \citep{iLiSAGithub} 
\item \texttt{Matplotlib}: A 2D graphics environment \citep{Hunter:2007} 
\item \texttt{Astropy}: a community-developed core Python package for Astronomy \citep{astropy:2013, astropy:2018}, 
\item \texttt{python-casacore}: A wrapper around CASACORE, the radio astronomy library \citep{python-casacoreGithub}
\item \texttt{KERN}: a bi-annually released set of radio astronomical software packages \citep{molenaar2018kern}
\item \texttt{pandas}: Python Data Analysis Library \citep{mckinney-proc-scipy-2010}
\item \texttt{NumPy}: the fundamental package for scientific computing with Python \citep{van2011numpy}
\item \texttt{SciPy}: Open source scientific tools for Python \citep{SciPy}
\item \texttt{H5Py}: HDF5 for Python \citep{H5Py,collette2013python}
\end{itemize}





\bibliographystyle{model2-names}
\bibliography{tester_paper}







\end{document}